\documentclass[mathematics,article,submit,moreauthors,pdftex]{Definitions/mdpi}
\usepackage{graphicx}
\usepackage{tabularx}
\usepackage{makecell}
\firstpage{1}
\pubvolume{1}
\issuenum{1}
\articlenumber{0}
\pubyear{2020}
\copyrightyear{2020}
\history{Received: date; Accepted: date; Published: date}

\pdfoutput=1

\Title{Bayesian analysis of population health data}

\Author{Dorota M\l{}ynarczyk$^1$*\orcidA{}, Carmen Armero$^2$\orcidB{}, Virgilio G\'omez-Rubio$^3$\orcidC{} and Pedro Puig$^{4,5}$\orcidD{}}

\AuthorNames{Dorota M\l{}ynarczyk, Carmen Armero, Virgilio G\'omez-Rubio and Pere Puig}

\address{%
$^{1}$ \quad Universitat Aut\`onoma de Barcelona; dorotaanna.mlynarczyk@uab.cat\\
$^{2}$ \quad Universitat de Val\`encia; carmen.armero@uv.es\\
$^{3}$ \quad Department of Mathematics, School of Industrial Egineeering-Albacete, Universidad de Castilla-La Mancha, Albacete (Spain); virgilio.gomez@uclm.es\\
$^{4}$ \quad Universitat Aut\`onoma de Barcelona; ppuig@mat.uab.cat\\
$^{5}$ \quad Centre de Recerca Matem\`atica (CRM), Universitat Aut\`onoma de Barcelona, Cerdanyola del Vall\`es, Barcelona, Spain}

\corres{Correspondence: dorotaanna.mlynarczyk@uab.cat; Departament de Matem\`atiques, Edifici C, 08193 Bellaterra (Cerdanyola del Vall\`es, Barcelona, Spain)}

\abstract{The analysis of population-wide datasets can provide insight on the health status of large populations so that public health officials can make data-driven decisions. The analysis of such datasets 
often requires highly parameterized models with different types of fixed and randoms effects to account for risk factors, spatial and temporal variations, multilevel effects and other sources on uncertainty.
To illustrate the potential of Bayesian hierarchical models, a dataset of about 500000 inhabitants released by the Polish National Health Fund containing information about ischemic stroke incidence for a 2-year period is analyzed using different types of models.
Spatial logistic regression and survival models are considered for analyzing the individual probabilities of stroke and the times to the occurrence of an ischemic stroke event.
Demographic and socioeconomic variables as well as drug prescription information are available at an individual level. Spatial variation is considered by means of region-level random effects.
}


\begin{document}
%%%%%%%%%%%%%%%%%%%%%%%%%%%%%%%%%%%%%%%%%%

%%%%%%%%%%%%%%%%%%%%%%%%%%%%%%%%%%%%%%%%%%

\section{Introduction}

Population and Public Health officials often require to address complex issues in important health problems with high levels of uncertainty
that can affect millions of people. Providing scientific evidence to help decision-making processes in that area is a key issue and statistical analysis  becomes an essential tool.

Data on large populations are often difficult to obtain due to confidentiality issues and the technical difficulties and financial resources involved in their design, maintenance and updating as well as its day-to-day management. The existence and availability of population databases for scientific exploitation is a treasure. Having a strong knowledge on the population makes it possible to accurately estimate the parameters of interest in the study, to identify potential risk factors, to detect patterns, outcomes or groups of individuals with special characteristics, and minimize the uncertainty associated with the prediction process. These studies are of great help to Public Health insofar as they contribute to the development of efficient and effective strategies and policies aimed at improving the health of the target population.

This paper deals with population health from a statistical point of view, and concentrates on the prevalence of stroke in Poland. In particular, we aim to identify different patterns that may increase the probabilities of suffering from a stroke. Stroke is one of the most serious diseases that can affect a person. It is the second most common cause of death globally, responsible for approximately $11\%$ of the world’s total deaths \cite{WHO}. Stroke often leads to permanent disability, which means partial or complete dependence on others and, consequently, to social withdrawal. It causes huge social costs related not only to the costs of hospital treatment, but most of all to long-term care and rehabilitation expenses as well as the inability to work with the necessity to pay a disability pension \cite{coststroke:2020}. Therefore, in order to improve prevention, various factors that may be associated with the occurrence of a stroke must be analyzed, which is what we do in this paper.

Prevention strategies primarily focus on eliminating or reducing the impact of modifiable risk factors and educating the entire society, in particular those predisposed to the disease. It is recommended to lead a healthy lifestyle on the basis of a regular physical activity, a balanced diet, and to stop smoking and drinking alcohol. Moreover, such actions also have a positive effect on the prevention of diseases such as diabetes and cancer \cite{prevention:2016}. Unfortunately, the risk of recurrent stroke increases every year and it is estimated at over $11\%$ at one year and at around $39\%$ at 10 years after initial stroke \cite{reccurence:2011}. Therefore, secondary prevention, including pharmacotherapy and rehabilitation, especially at long-term, is very important.

In Poland, knowledge about stroke is still insufficient, but there are educational activities and social campaigns that will hopefully be effective in the future \cite{IOZ}. In 2019, the Polish National Health Fund released an anonymized dataset about 500$\hspace*{0.04cm}$000 inhabitants that included information about ischemic stroke and other important covariates such as gender, age, administrative region and drug prescriptions. Data   from this paper come from that study and they were made available for the Digital Health Hackathon- Forum eHealth in 2019 \cite{Dataset}.

Spatial logistic regression is an appropriate statistical procedure for estimating the probability of suffering from a stroke with regard to demographic and socioeconomic characteristics of the individuals as well as their pharmacological treatment administered \cite{BivandGomezRubio:2021}. This model also includes spatial random effects that account for the regional variation of the incidence of stroke. As in our case, when the database includes not only whether or not each individual has had a stroke, but also the exact date of the event for those who have had experienced it, the problem can be recasted as a time-to-event analysis for which survival models can be used \cite{Ibrahim:2005}. Similarly, spatial frailties can subsequently be employed to account for regional variations.

Bayesian statistics provides a suitable inference
 on the different unknown elements of the model and their uncertainty. Given the dimension
 of the dataset, typical computational methods for model fitting based on
Markov chain Monte Carlo (MCMC) procedures \cite{Brooksetal:2011} may not be adequate.
For this reason, the integrated nested Laplace approximation (INLA)
\cite{Rueetal:2009} will be used to estimate the marginal posterior distribution of the
model parameters and other quantities of interest.

This paper is organized as follows. Section~\ref{sec:models} introduces the statistical models used in this paper. Logistic and survival regression are presented in Subsections \ref{sec:logit} and \ref{sec:survival}, respectively, and a short introduction to the integrated nested Laplace approximation (INLA) is included in Subsection \ref{sec:INLA} within the framework of Bayesian inference. Section~\ref{sec:example} is devoted to the study on the stroke and associated risk factors in Poland, where the Polish stroke dataset is explored. Finally, Section~\ref{sec:discussion} includes a summary of the results and a final discussion.

\section{Statistical models}
\label{sec:models}

Regression and survival methods are usually relevant procedures in population studies concerning diseases and associated risk factors. In these cases, the outcomes of interest tend to focus on the study of the prevalence of the disease in a given time period and the length of time until its occurrence. The estimation of the probability associated to the disease in terms of a set of explanatory covariates and random effects is often modeled using mixed logistic regression models \cite{Christensen:2011}. Survival models are statistical models specially designed to learn about time-to-event outcomes and their relationships with regard to relevant risk factors \cite{Ibrahim:2005}. They also include as a particular issue the assessment of prevalence probabilities by means of particular cases of the survival function. Both approaches and how they are related to each other are described below in a first subsection devoted to logistic regression and a second one for survival models. This section concludes with an introduction to the integrated nested Laplace approximation (INLA) \cite{Rueetal:2009} within the framework of Bayesian inferential methods.

\subsection{Logistic regression}
\label{sec:logit}

Binomial regression connects probabilities associated to Bernoulli trials with  covariates. The outcome of interest is an observable binary response which describes the presence (value 1) or absence (value zero) of a certain individual feature of the   population under study. In the case of individual $i$ it is defined as follows

$$
O_i \sim Ber(p_i),
$$
\noindent
being $p_i$ the probability of success in the subsequent Bernoulli trial. Probabilities and covariates are not usually in the same scale. For this reason a link function $g$ is defined in order to accommodate the probabilities and the linear predictor $\eta_i$ in the same scale as follows

 \begin{equation}
 g(p_i)=\eta_i= \beta_0+\beta_1 x_{i1}+\ldots +\beta_q x_{iq},
 \label{eqn:regression1}
 \end{equation}

\noindent where $\boldsymbol \beta=(\beta_0, \beta_1, \ldots, \beta_q)^{\prime}$ is the regression coefficient vector associated to covariates $\boldsymbol x_i=(x_{i0}=1,x_{i1}, \ldots, x_{iq})^{\prime}$. The most common link functions when dealing with binary variables are the logit and the probit functions. The logit function is the canonical link function for the Bernoulli distribution in generalized linear models and a binomial regression endowed with the logit link function is called logistic regression. It offers an intuitive interpretation of the relationship between the probability of interest and the linear predictor in terms of odds in logarithmic scale as follows $$ \eta_i=\mbox{logit}(p_{i})= \mbox{log}\Big (\frac{p_i}{1-p_i} \Big).$$

Random effects allow to assess variability associated to the outcome of interest within groups that are not accounted for the covariates. Random effects can be modeled in different ways. The simplest one considers random effects as conditionally independent and identically distributed random variables
with Gaussian distribution of zero mean and precision (\textit{i.e.}, the reciprocal of the variance)  $\tau$.
This condition assumes that given $\tau$ there is no
prior correlation among the different groups and that differences among them
are only due to intrinsic factors. The inclusion of these elements in the regression model forces its reformulation with the addition of a new index to indicate the random effect associated to group $j$, $j=1,\ldots, J$, as follows:
 \begin{align}
 O_{ij} & \sim Ber(p_{ij}),\nonumber\\
 & \mbox{logit}(p_{ij})=\eta_{ij}= \boldsymbol x_{ij}^{\prime} \boldsymbol \beta + \boldsymbol \gamma_{j},
 \label{eqn:regression2}
 \end{align}
\noindent where $\gamma_{j} \mid \tau \sim \mbox{N}(0, \,\tau)$. It is worth noting that the model can include covariates associated with groups. In such scenarios, the value of the corresponding covariate would be the same for all individuals belonging to the same group $j$.

In the case where it is assumed that the risk varies smoothly along the study region, spatially correlated random effects can be considered. A typical approach considers the Intrinsic Conditional Auto-Regressive (ICAR) model \cite{Besag74} that incorporates information from the neighboring regions. This model specifies a Gaussian distribution for the conditional distribution of the random effect $\gamma_j$ associated to the region $j$, $j=1\ldots,J$
given the set of the random effects at its neighbours (denoted by $l\sim j)$ with mean $\sum_{l\sim j} \, \gamma_l/ n_j$ and precision $\tau / n_j$, where $n_j$ is the number of neighbours of region $j$. This model is often used in disease mapping models to account for spatial and spatio-temporal risk variation. The joint distribution for $\boldsymbol \gamma=(\gamma_1, \ldots, \gamma_J)^\prime$ is a multivariate normal random vector
\begin{equation}
\boldsymbol \gamma \sim \mbox{N}(\boldsymbol 0, Q ),
\end{equation}
 \noindent where $Q$ is a $J \times J$
precision matrix with entries $n_j, \,j=1\ldots,J$ in the diagonal and entries $Q_{jl}$ equal to -1 if regions $j$ and $l$ are neighbours and 0 otherwise.
Given that this is an improper distribution, a sum-to-zero constraint is often
added on the values of the random effects, i.e., $\sum_{j=1}^J \gamma_j = 0$
\cite{Banerjeeetal:2014}.

Leroux \textit{et al.} (1999) \cite{Lerouxetal:1999} propose an alternative specification for the precision matrix that better distinguishes between spatial dependence and overdispersion effects as follows:

$$
 (1-\phi) I + \phi Q,
$$
where $I$ is the identity matrix and parameter $\phi \in [0,\,1]$ determines how matrices $I$ and $Q$ are combined. Values of $\phi$ close to 0 indicate that there is a weak spatial pattern, while values close to 1 mean a strong spatial pattern. 

\subsection{Accelerated failure time survival models }
\label{sec:survival}

Survival analysis is the branch of Statistics  dedicated to the study of the length of time between two events, the event that initiates the observation process and the final event, also called the event of interest  or final point,   which determines the end of the monitoring procedure. From a statistical point of view, the topic focuses on the analysis of samples from  random variables with support in the positive real numbers, generally skewed and usually partially observed.  In most cases the observation period ends before the event of interest occurs and the actual observation period does not always coincide with its theoretical start.  In the first case, the data will be right censored and left truncated in the second one.  Both mechanisms, especially censoring,   introduce complexity into the statistical analysis due to their  important role in the likelihood function.

The key concepts for assessing survival times are the survival and the hazard function. The survival function for the survival random variable $T_i$ at $t \geq 0$ corresponding to individual $i$ is the probability that this individual  survives beyond   time   $t$ as defined below

\begin{equation} \label{eqn:def_survivalfunction}
S_i(t )=P(T_i \geq t ).
\end{equation}

The hazard function of $T_i$ at  time $t$ is a nonnegative function that describes the instantaneous rate of  occurrence of the event  among  individuals who have not yet experienced the event of interest
 at  $t$.   It is defined in terms of a conditional probability as  follows:

  \begin{equation} \label{eqn:def_hazardfunction}
 h_i(t) = \lim_{\Delta t \to 0} \frac{P(t \leq T_i < t+\Delta t \mid T_i \geq t )}{\Delta t}.
\end{equation}
The  hazard function is very popular in epidemiological contexts  where  it is known as the incidence function.

Survival regression models assess the variability of the survival times of the different individuals of the target population with regard to relevant  covariates. Accelerated failure time (AFT) models  are, together with Cox proportional hazards models, the most popular in survival analysis \cite{Ibrahim:2005}. We start assuming a basic   AFT model for the survival time of individual $i$ as follows

\begin{equation}
\label{eqn:aft}\mbox{log}\,(T_i)= \boldsymbol x_i^{\prime} \boldsymbol \beta + \sigma\,\epsilon_i,
\end{equation}
\noindent being  $\boldsymbol x_i$ and $\boldsymbol \beta$ the same as in (\ref{eqn:regression1}), $\sigma$ a scale parameter and $\epsilon_i$ $i.i.d$ random variables with a
standard Gumbel distribution (standard  type I Fisher-Tippett extreme value distribution). This is a non-negative continuous distribution with probability density function
   $f_i(t)=\mbox{e}^{t}\,\mbox{exp}\{-\mbox{e}^t\}$, survival function $S_i(t)= \mbox{exp}\{-\mbox{e}^t\}$, and hazard function $h_i(t)=\mbox{e}^{t}$, $t>0$.
 As a result, the distribution of $T_i$ is a Weibull distribution with
shape parameter $1/\sigma$ and scale parameter $\exp\{-\boldsymbol x_i^{\prime} \boldsymbol \beta / \sigma\}$, i.e., it has hazard function

 \begin{equation}
 \label{eqn:Weibull}
h_i(t) = \exp\{ - \boldsymbol x_i^{\prime} \boldsymbol \beta /\sigma\}\,\frac{1}{\sigma}\,t^{\frac{1}{\sigma} - 1}.
\end{equation}

   The AFT model in (\ref{eqn:aft})   is very flexible because it can also be expressed as a Cox  proportional hazards model \cite{Kalbfleisch:1980,Cox:1984}.

As in the binomial regression model, the inclusion of random effects associated with groups of individuals in the survival model also needs a new definition format. Assuming the same type of random effects $\gamma_j$ that we have considered in the logistic regression model, our accelerated model will be as follows:
\begin{equation}
\label{eqn:aft2}
\mbox{log}\,(T_{ij})= \boldsymbol x_{ij}^{\prime} \boldsymbol \beta + \gamma_j+ \sigma\,\epsilon_{ij},
\end{equation}
\noindent with the   $\gamma_j$'s modelled according to each of the two proposals, conditionally \textit{i.i.d.} and spatially correlated,  formulated as in the previous sub-section. 

\subsection{Bayesian inference and the integrated nested Laplace approximation}
\label{sec:INLA}

Bayesian inference accounts for  uncertainty in terms of probability distributions. The main element of a Bayesian learning process is the likelihood function, which is constructed from the sampling model and the observed data $\mathcal D$, and the prior distribution for all unknown elements in the sampling model. The subsequent posterior distribution combines two pieces of information and is computed via  Bayes' theorem.

Inference
for hierarchical and highly parameterized models is often
conducted using a number of tools available. Markov chain Monte Carlo (MCMC)
methods can estimate a wide range of models but they are
  too slow when dealing with large datasets such as those arising from population studies \cite{Bardenetetal:2017}.

Alternatively, approximate inference could be carried out so that posterior sampling is not
required. In particular, the integrated nested Laplace approximation (INLA) \cite{Rueetal:2009} provides accurate approximations of the posterior marginal
distribution for the latent effects, parameters and  hyperparameters of the model.  INLA  considers random samples  from a common probabilistic population as conditionally independent given a latent Gaussian Markov random field
(GMRF) \cite{RueHeld:2005} $\boldsymbol \theta$ with zero mean and precision matrix $H$ that depends on some hyperparameters $\boldsymbol \phi$ which can include effects of different
type (regression coefficients, random effects, seasonal effects, etc).   This feature ensures
that the structure of $H$ is sparse so that computationally efficient
algorithms can be employed for the estimation procedure.

  The hierarchical Bayesian model stated by INLA can be generally formulated as

$$
\pi(\boldsymbol \theta, \boldsymbol \phi  \mid \mathcal{D}) \propto
\mathcal L(\boldsymbol \theta, \boldsymbol \phi)
  \pi(\boldsymbol \theta \mid \boldsymbol \phi) \pi(\boldsymbol \phi),
$$
\noindent
where  $\pi(\boldsymbol \theta, \boldsymbol \phi  \mid \mathcal{D}) $ is the posterior distribution of $(\boldsymbol \theta, \boldsymbol \phi)$,
$\mathcal L(\boldsymbol \theta, \boldsymbol \phi)$ represents the likelihood function of $(\boldsymbol \theta, \boldsymbol \phi)$ for data $\mathcal D$,   $\pi( \boldsymbol \theta \mid   \boldsymbol \phi )$   is the conditional GMRF discussed above and
$\pi(\boldsymbol \phi)$ is the prior distribution for   hyperparameters $\boldsymbol \phi$.

INLA starts  the estimation procedure  by obtaining a good approximation to the joint posterior distribution of the hyperparameters, i.e.,  $\pi(\boldsymbol \phi \mid \mathcal D)$.
Then it uses this approximation to compute the posterior marginal of  each univariate hyperparameter
 $\phi_l$ and the marginal posterior distribution of each latent term $\theta_m$ in $\boldsymbol \theta$  as follows

$$
\pi(\phi_l \mid \mathcal D) = \int \pi(\boldsymbol \phi \mid \mathcal D)\, d\boldsymbol \phi_{-l},
$$

$$
\pi(\theta_{m} \mid \mathcal D) \propto \int \pi(\theta_{m} \mid \boldsymbol \phi, \mathcal D) \,\pi(\boldsymbol \phi \mid \mathcal D) \,d\boldsymbol \phi.
$$
\noindent
These integrals are approximated using numerical integration methods and the
Laplace approximation \cite{Rueetal:2009,GomezRubio:2020}.

 Note that
once the posterior marginals are available it is possible to compute
quantities of interest about the parameters and hyperparameters such as
  posterior means or credible intervals.

The INLA procedure  is implemented in the R-INLA package \cite{INLA} for the R statistical software \cite{R}.
This package
  can also be used to compute a number of features for model selection, which
include information-based criteria such as the deviance information criterion
\citep[DIC,][]{spiegelhalteretal:2002} and the Watanabe-Akaike information criterion \citep[WAIC,][]{Watanabe:2013}.

\section{Analysis of ischemic stroke and risk factors in Poland}
\label{sec:example}

In Poland, the incidence of stroke is similar to that in other European countries: approximately 112 strokes per 100$\hspace*{0.04cm}$000 inhabitants, which gives about 65$\hspace*{0.04cm}$000 new cases of stroke registered annually \cite{SAFE}. The number of strokes in Poland is expected to increase in the coming years, what is mostly related to the aging of the population. This means an increased demand for medical and palliative care, which require both adequate resources and the development of a strategy for the future \cite{IOZ}.

As presented in the introduction, the data for the study consist on an anonymized dataset of about 500$\hspace*{0.04cm}$000 inhabitants from the Polish National
Health Fund that includes individual information about ischemic stroke and other important covariates such as gender, age, administrative region and drug prescriptions. The period of observation is two years, but the actual dates have not been released and they remain unknown.  We do not know the reasons for this decision; we can only assume that it is a recent period of two years. The patient's age is given in 5-years-old groups and the gender is a binary variable without clearly indication of which value stands for which gender. However, it is commonly known that women live longer than men and thus we can distinguish the two genders in the data. We decided to analyze only patients older than 38 years old, as in younger age groups stroke had a very low prevalence. As a result, the three age groups finally considered in the analysis are (38-58] years (group Age1), (58, 68] (group Age2), and (68, 108] (group Age3). As we are interested in studying spatial dependencies, we take only patients with known territorial code (no missing values). The final dataset consists on 332$\hspace*{0.04cm}$799 patients, among them 2$\hspace*{0.04cm}$889 had ischemic stroke $(0.9\%)$. This percentage is low, but due to the fact that the sample is probably randomly selected (they are not people with a specific disease or medical history) and the observation period lasted only two years, it seems reasonable.

For almost each patient, territorial code of the   place of registration is available. Based on that code, we can classify to which powiat or voivodeship belongs the patient. In Poland, powiat is a second-level local government unit, which is often referred to as 'county', and it is a part of a larger unit-voivodeship.
In the dataset, there are 379 powiat-level entities, which can be divided according to the administrative divisions of Poland into 66 city counties (formally 'Cities with powiat rights') and 313 regular counties, which we will be called land counties. Nowadays there are 380 powiats, which have changed in 2013 and therefore we assume that the dataset comes from two consecutive years between 2003 and 2013 \cite{DzU:821,DzU:853}. Poland is divided in 16 voivodeships, which could also be used instead of county divisions.

The dataset also contains information on prescriptions for reimbursed drugs. For each prescription the three-digits code of the Anatomical Therapeutic Chemical (ATC) Classification System is provided. Based on this code, the drug can be identified on which organ or system it acts. In this classification, there are five different levels to identify the active substances of any drug. In the dataset, the three-digits codes allow to classify the prescription in a pharmacological or therapeutic subgroup. Hence we decided to include also the information of the prescriptions dispensed by patient. The risk factors for stroke are, among others, high blood pressure, atrial fibrillation (AF) and diabetes \cite{Boehme:2017}. Therefore we choose to include in the analyses the use of prescriptions for the cardiovascular system (based on the ATC classification - type 'C'), any antithrombotic agents (used in the prevention or treatment of AF, ATC B01) and drugs used in diabetes (ATC A10), because they appear to be the most relevant when analyzing the occurrence of strokes \cite{WHO:2020}. In our analysis, it is not possible to detect any association between the stroke and the prescription drug, and its associated disease. This should be beared in mind when interpreting the results, i.e., the coefficients associated to this covariates will assess the relation  of suffering from the condition and taking the associated prescription drugs.

The impact of  socioeconomic factors  cannot be overlooked when talking about such a complex disease as stroke. People with a lower status have limited access to medical care, which may result in the lack of quick diagnosis, which in the event of a stroke may lead to severe disability. Low level of public awareness can be related with the increase of risk factors for stroke and can affect recovery during rehabilitation. This is consistent with studies showing that low socioeconomic status may result in an increased incidence of stroke and mortality \cite{Addo:2012}.
Accordingly, we included in the study the powiat index of deprivation (PID). This index is computed from five components using data from 2013  from another database independent of the one used in our study.\cite{Smetkowski:2015}: income, employment, living conditions, education and access to goods and services. The values of the index are in the range of $-1.8$ to $+1.1$, with a negatively skewed distribution (with zero mean and standard deviation 0.58). A higher value of the index means a higher risk of deprivation to which the population of a given powiat is exposed.

In the final dataset there were less than $1\%$ patients who suffered a stroke. Almost half of the population is over 38 and under 58, and more than half are women and people living in land counties. The vast majority of patients takes drugs for the cardiovascular system, while drugs for diabetes and atrial fibrillation (and others) represent only around $12\%$. Table \ref{tab: sum_descr} shows a short description of the percentage of people who have  and have not suffered a stroke with regard to age, gender, county type and group of medicines.

\begin{table}[ht]
\caption{Summary statistics of the dataset (\%)}
\centering
\resizebox{\textwidth}{!}{%
\label{tab: sum_descr}
\begin{tabular}{c|cc|cc|cc|cc|cc|cc}
 \hline
AGE GROUP & \multicolumn{2}{c|}{STROKE} & \multicolumn{2}{c|}{GENDER} & \multicolumn{2}{c|}{COUNTY TYPE} & \multicolumn{2}{c|}{ATC C} & \multicolumn{2}{c|}{ATC A10} & \multicolumn{2}{c}{ATC B01}\\
 \hline
 & NO & YES & MEN & WOMEN & LAND & CITY & NO & YES & NO & YES & NO & YES \\
 \hline
Age1 (38-58] & 46.76 & 0.13 & 22.07 & 24.81 & 31.67 & 15.22 & 31.94 & 14.95 & 44.53 & 2.36 & 43.89 & 3.00 \\
Age2 (58-68] & 26.69 & 0.22 & 12.11 & 14.81 & 17.21 & 9.71 & 8.72 & 18.20 & 22.39 & 4.52 & 23.75 & 3.16 \\
Age3 (68-108] & 25.68 & 0.51 & 9.62 & 16.58 & 16.31 & 9.89 & 3.74 & 22.45 & 19.66 & 6.54 & 20.82 & 5.37 \\
 \hline
TOTAL & 99.13	&0.86	&43.80	&56.20&	65.19&	34.82&	44.40	&55.6	&86.58	&13.42	&88.46	&11.53\\
\hline
\end{tabular}}
\end{table}

\subsection{Bayesian logistic and survival modelling}

Let $p_{ij}$ be the probability that the individual $i$ living in powiat $j$ will suffer an ischemic stroke, and $T_{ij}$ be the time when that individual suffers a stroke since entering the study. The statistical analysis begins with a basic logistic regression and a basic accelerated failure time survival model for analyzing the probability $p_{ij}$ and the survival time $T_{ij}$, respectively, in terms of covariates gender, age, prescriptions for reimbursed drugs, and PID as follows:
\begin{align}
\label{eqn:powiatmodels}
\mbox{logit} (p_{ij}) & = \boldsymbol x_{ij}^{\prime} \,\boldsymbol \beta \nonumber \\
\mbox{log} (T_{ij}) & =\boldsymbol x_{ij}^{\prime} \,\boldsymbol \beta + \sigma \epsilon_{ij} \nonumber \\
 \boldsymbol x_{ij}^{\prime} \,\boldsymbol \beta &= \beta_0+\beta_1 I_{Woman}(ij)+\beta_2 I_{Age2}(ij)+\beta_3 I_{Age3}(ij)+ \beta_4 I_{City}(ij)+ \nonumber \\ &+\beta_5 I_{T.A10}(ij) +\beta_6 I_{T.B01} (ij)+\beta_7 I_{T.C}(ij)(ij) + \beta_8 Depr(j),
\end{align}

\noindent where $I_A(ij)$ is an indicator variable for $A$ that takes the value 1 if the individual $i$ from powiat $j$ has the characteristic $A$ and zero if she or he does not, and consequently
 $I_{Woman}(ij)$, $I_{Age2}(ij)$, $I_{Age3}(ij)$, $I_{City}(ij)$, $I_{T.A10}(ij)$, $I_{T.B01}(ij)$ and $I_{T.C}(ij)$ are the indicator random variables for being a woman, being in age group Age2, Age3, living in city county and having received diabetes, antithrombotic and cardiovascular treatment in powiat $j$, respectively. The $Depr$ covariate stands for the deprivation index which is the numerical variable defined for each powiat. To complete the specification of the Bayesian model it is necessary to elicit a prior distribution for the parameters and hyperparameters of the model. In the case of the logistic regression model the set of parameters $\boldsymbol \theta=(\beta_0, \beta_1, \ldots, \beta_8)^{\prime}$ is a GMRF with diagonal precision matrix 0.001 for all the coefficients except for $\beta_0$ whose marginal prior distribution is selected as an improper  
Gaussian distribution with zero mean and zero precision.

The discussion of   the marginal prior distribution for the scale parameter $ \sigma$ in the survival model needs a previous comment about INLA and the Weibull distribution.
INLA offers two different parameterizations of the Weibull
distribution for survival models. We have opted for the so-called first variant, which
corresponds to shape parameter $\alpha = 1 / \sigma$ and scale parameter $\lambda = \exp\{\boldsymbol x_{ij}^{\prime} \boldsymbol \zeta\}$, so that the
hazard function of $T_{ij}$ is

$$
h_{ij}(t) = \lambda \alpha t^{\alpha -1} = \exp\{\boldsymbol x_{ij}^{\prime} \boldsymbol \zeta\} \alpha t^{\alpha - 1}.
$$

This parameterization implies that positive coefficients $\zeta$'s of the covariates increase hazard, while negative values reduce it. Note that this
parameterization is slightly different from the typical parameterization of this AFT model shown in equation (\ref{eqn:Weibull}). Coefficients $\zeta$'s estimated with INLA are equal to coefficients $-\beta/\sigma$'s in the accelerated survival model \cite{Wangetal:2018}.

 The shape parameter $\alpha$ of the Weibull distribution has a penalized complexity prior (PC-prior) \cite{Simpsonetal:2017}. In fact, INLA considers
$\alpha = \exp\{0.1\alpha^{\prime}\}$ to avoid numerical instabilities and the prior is set on $\alpha^{\prime}$. PC-priors are defined using the Kullback-Leibler distance between the proposed model and a natural base model, which in this
case corresponds to $\alpha = 1$, that is the exponential distribution. In our model, we have used the default
PC-prior for $\alpha$; see \cite{vanniekerk2020principled} for details.

 Random effects associated to the powiats are introduced in the logistic and the survival model in (\ref{eqn:powiatmodels}) according to the two proposals presented in the previous section: in terms of conditionally $i.i.d.$ random variables and spatially correlated random variables. The marginal prior distribution of the precision $\tau$ in the case of both conditionally independent and spatially correlated random effects is an improper uniform distribution in the interval 0 to infinity. The weight parameter $\phi$ in the precision
of the spatial effect has a prior distribution so that the logit of $\phi$ follows a
Gaussian distribution with zero mean and precision 0.1.

Table \ref{tab: results} presents the posterior mean and posterior credible intervals for the parameters and hyperparameters of the logistic regression model and the accelerated survival model without random effects, and with random effects in terms of conditionally $i.i.d$ random variables and spatially correlated random variables. Moreover all models have been evaluated through the  DIC and  WAIC criteria. For both types of modelling (i.e., logistic and survival), the model with spatially correlated random effects has the lowest values of DIC and WAIC.

\begin{table}[ht]
\centering
\caption{Posterior summaries for the parameters and hyperparameter of the logistic regression model and survival model without random effects (LOGIT and SURVIVAL ), with random effects in terms of conditionally $i.i.d$ random variables (LOGIT IID and SURVIVAL IID ), and spatially correlated random variables (LOGIT SPATIAL and SURVIVAL SPATIAL). }
\resizebox{\textwidth}{!}{%
\label{tab: results}
\begin{tabular}{ll|c|c|c|c|c|c}
 \hline
covariable & & LOGIT & SURVIVAL & LOGIT IID & SURVIVAL IID & LOGIT SPATIAL & SURVIVAL SPATIAL \\
 \hline
intercept & \makecell{ mean \\ CI } & \makecell{-5.901\\ (-6.013, -5.791)} & \makecell{-5.902\\ (-6.014, -5.793)} & \makecell{-5.925\\ (-6.042, -5.811)} & \makecell{-5.925\\ (-6.041, -5.811)} & \makecell{-5.912\\ (-6.061, -5.764)} & \makecell{-5.914\\ (-6.072, -5.757)} \\
 Woman & \makecell{ mean \\ CI } & \makecell{-0.217\\ (-0.291, -0.142)} & \makecell{-0.214\\ (-0.288, -0.14)} & \makecell{-0.217\\ (-0.291, -0.142)} & \makecell{-0.214\\ (-0.288, -0.14)} & \makecell{-0.216\\ (-0.291, -0.142)} & \makecell{-0.214\\ (-0.288, -0.14)} \\
 Group Age2 (58-68] & \makecell{ mean \\ CI } & \makecell{0.935\\ (0.812, 1.058)} & \makecell{0.933\\ (0.81, 1.056)} & \makecell{0.933\\ (0.81, 1.057)} & \makecell{0.931\\ (0.809, 1.055)} & \makecell{0.933\\ (0.81, 1.057)} & \makecell{0.932\\ (0.809, 1.055)} \\
 Group Age3 (68-108]& \makecell{ mean \\ CI } & \makecell{1.729\\ (1.613, 1.847)} & \makecell{1.722\\ (1.606, 1.84)} & \makecell{1.729\\ (1.612, 1.847)} & \makecell{1.722\\ (1.605, 1.839)} & \makecell{1.728\\ (1.611, 1.846)} & \makecell{1.72\\ (1.604, 1.838)} \\
 City county & \makecell{ mean \\ CI } & \makecell{0.122\\ (-0.015, 0.258)} & \makecell{0.121\\ (-0.015, 0.256)} & \makecell{0.07\\ (-0.104, 0.242)} & \makecell{0.07\\ (-0.1, 0.24)} & \makecell{0.007\\ (-0.17, 0.184)} & \makecell{0.007\\ (-0.173, 0.183)} \\
 T.A10 & \makecell{ mean \\ CI } & \makecell{0.238\\ (0.149, 0.326)} & \makecell{0.235\\ (0.147, 0.322)} & \makecell{0.238\\ (0.149, 0.326)} & \makecell{0.235\\ (0.147, 0.322)} & \makecell{0.239\\ (0.15, 0.327)} & \makecell{0.236\\ (0.148, 0.324)} \\
 T.B01 & \makecell{ mean \\ CI } & \makecell{0.235\\ (0.141, 0.328)} & \makecell{0.234\\ (0.14, 0.326)} & \makecell{0.236\\ (0.141, 0.329)} & \makecell{0.234\\ (0.14, 0.326)} & \makecell{0.236\\ (0.141, 0.329)} & \makecell{0.234\\ (0.14, 0.326)} \\
 T.C & \makecell{ mean \\ CI } & \makecell{0.324\\ (0.224, 0.425)} & \makecell{0.322\\ (0.222, 0.423)} & \makecell{0.325\\ (0.224, 0.426)} & \makecell{0.323\\ (0.223, 0.424)} & \makecell{0.324\\ (0.224, 0.425)} & \makecell{0.322\\ (0.222, 0.423)} \\
 Deprivation index & \makecell{ mean \\ CI } & \makecell{0.179\\ (0.092, 0.265)} & \makecell{0.178\\ (0.091, 0.263)} & \makecell{0.128\\ (0.012, 0.243)} & \makecell{0.129\\ (0.015, 0.242)} & \makecell{0.096\\ (-0.022, 0.214)} & \makecell{0.095\\ (-0.025, 0.213)} \\
 \hline
precision $\tau$ & \makecell{ mean \\ CI } & & & \makecell{16.833\\ (11.063, 22.566)} & \makecell{18.562\\ (13.66, 25.462)} & \makecell{11.306\\ (8.472, 15.486)} & \makecell{9.365\\ (4.78, 14.554)} \\
 \hline
shape parameter $1/\sigma$ & \makecell{ mean \\ CI } & & \makecell{1.114\\ (1.075, 1.155)} & & \makecell{1.112\\ (1.08, 1.149)} & & \makecell{1.112\\ (1.079, 1.147)} \\
 \hline
parameter $\phi$ & \makecell{ mean \\ CI } & & & & & \makecell{0.866\\ (0.746, 0.925)} & \makecell{0.889\\ (0.747, 0.978)} \\
 \hline
DIC & \makecell{ mean \\ CI } & 31296.11 & 31263.72 & 31258.58 & 31225.10 & 31231.96 & 31200.00 \\
WAIC & \makecell{ mean \\ CI } & 31296.14 & 31263.49 & 31256.18 & 31222.83 & 31230.30 & 31198.86 \\
 \hline
\end{tabular}}
\end{table}

All models have similar estimates of the regression coefficients associated with the covariates, providing evidence of   statistical robustness. As expected, a lower risk of stroke is associated with being a woman and age increases the risk of stroke. Naively, this can be regarded as if the results pointed to that being male increases the stroke rate by about 25$\%$ and being in the older age group multiplies the stroke rate by about 5-6 times.   The estimates of the model indicate that men older than 68 who live in a city county  have the highest risk of stroke. It is worth noting the positive relation between the pharmacological prescriptions dispensed to patients and the risk of stroke, specially those related to the cardiovascular system. The analysis of the credible intervals suggests that all the covariates, except the county, are relevant both for the risk of stroke and for time to  stroke. The risk of stroke grows in proportion to the deprivation index although the importance of this variable is questionable.
The posterior mean of the parameter $1/\sigma$ of the survival models is always close to 1. It could suggests that the risk of stroke does increase with time, but not rapidly. This latter may be due to the fact that the data was collected only for a period of two years and relates to people without specific diseases.

The posterior mean of the hyperparameter $\phi$ which assesses the strength of the spatial effect in the spatial models is equal to 0.866 and 0.889 for the logistic regression  and for the survival model, respectively, with 95$\%$ credible intervals that clearly state the relevance of the spatial effect. The posterior mean of the precision $\tau$ estimated for counties indicates that there is variation between powiats. It is lower for the spatial models, but still this is an evidence of the dispersion in the data.

\begin{figure}
 \centering
\includegraphics[width=\textwidth]{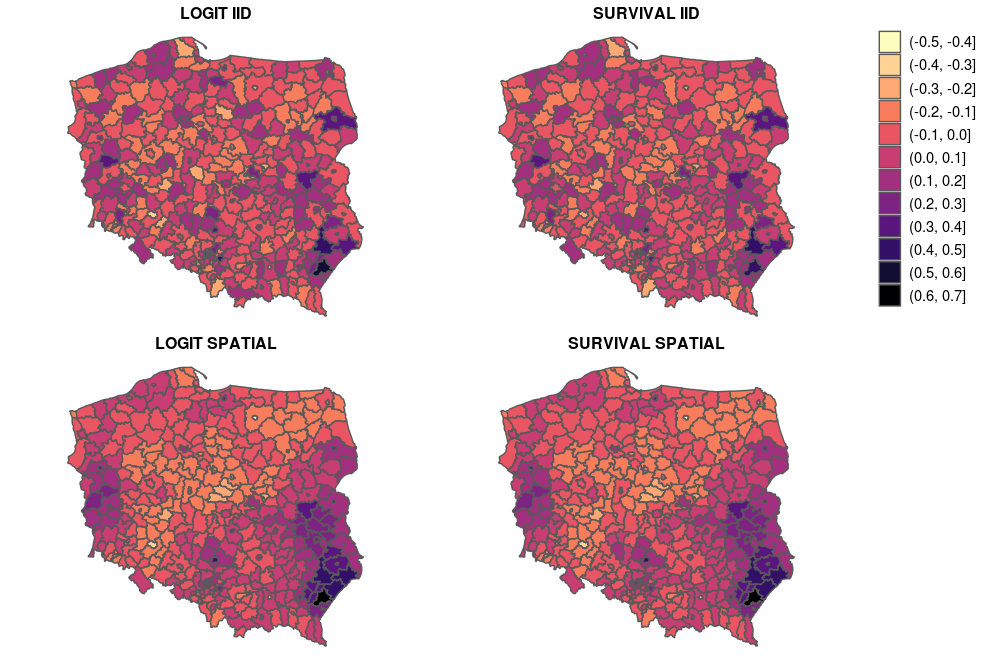}
\caption{Posterior mean for the conditional $i.i.d$ random variables in   LOGIT IID and SURVIVAL IID models, and for spatially correlated random variables in LOGIT SPATIAL and SURVIVAL models.  }
\label{fig:random_effects}
\end{figure}

Figure \ref{fig:random_effects} illustrates the posterior mean of the random effects for both the logistic regression and the survival modelling. As expected, the outcomes associated to the different powiats in the conditional $i.i.d$ models are very similar as well as those for the two spatial effects models. There are, however, differences between the conditional  $i.i.d$  and spatial models. The latter show strong spatial patterns, with an southwest-northeast alignment of the smallest values, which can be interpreted as regions with lower probability of stroke. On the contrary, a high-value cluster in the southeast, means that the risk of stroke is higher than in the other parts of the country.

\begin{figure}[hbt!]
 \centering
\includegraphics[width=\textwidth, height=15cm]{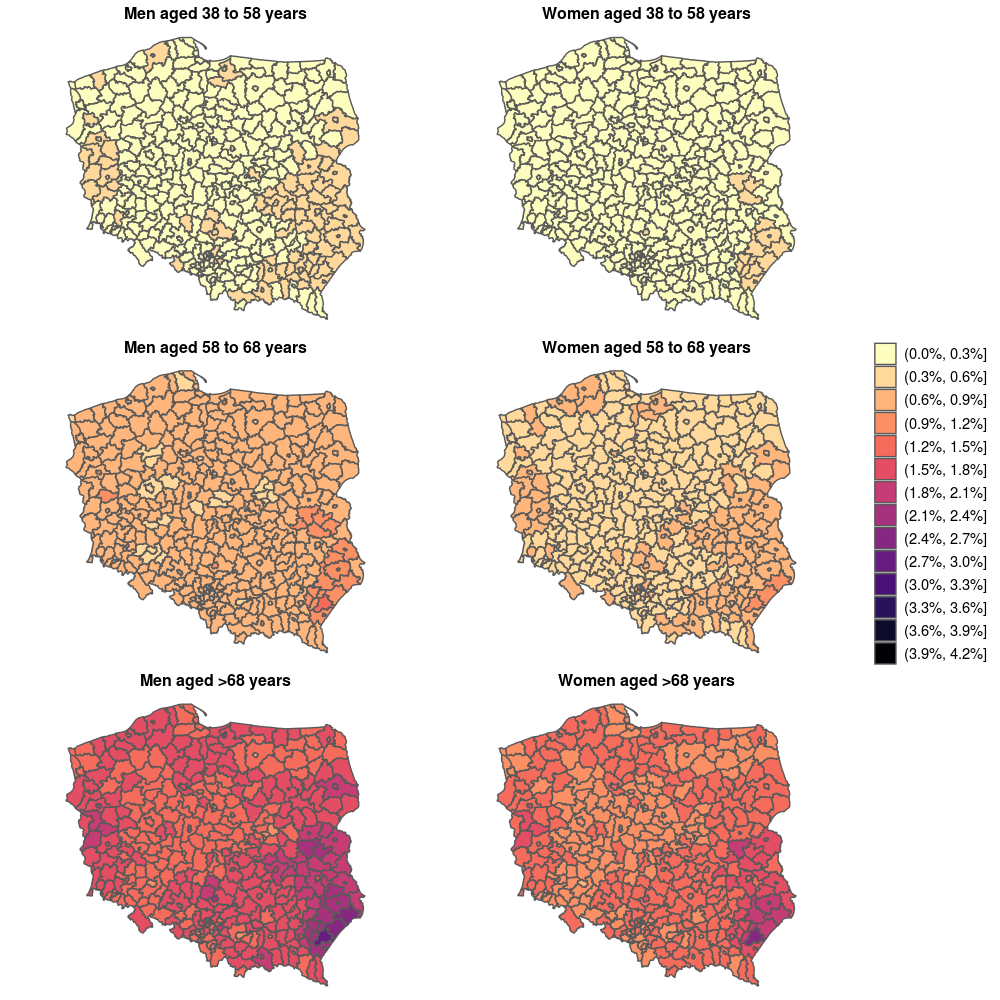}
\caption{Estimated probability of stroke by gender and age group based on the LOGIT SPATIAL model.}
\label{fig:results}
\end{figure}

\begin{figure}[hbt!]
 \centering
\includegraphics[width=\textwidth, height=15cm]{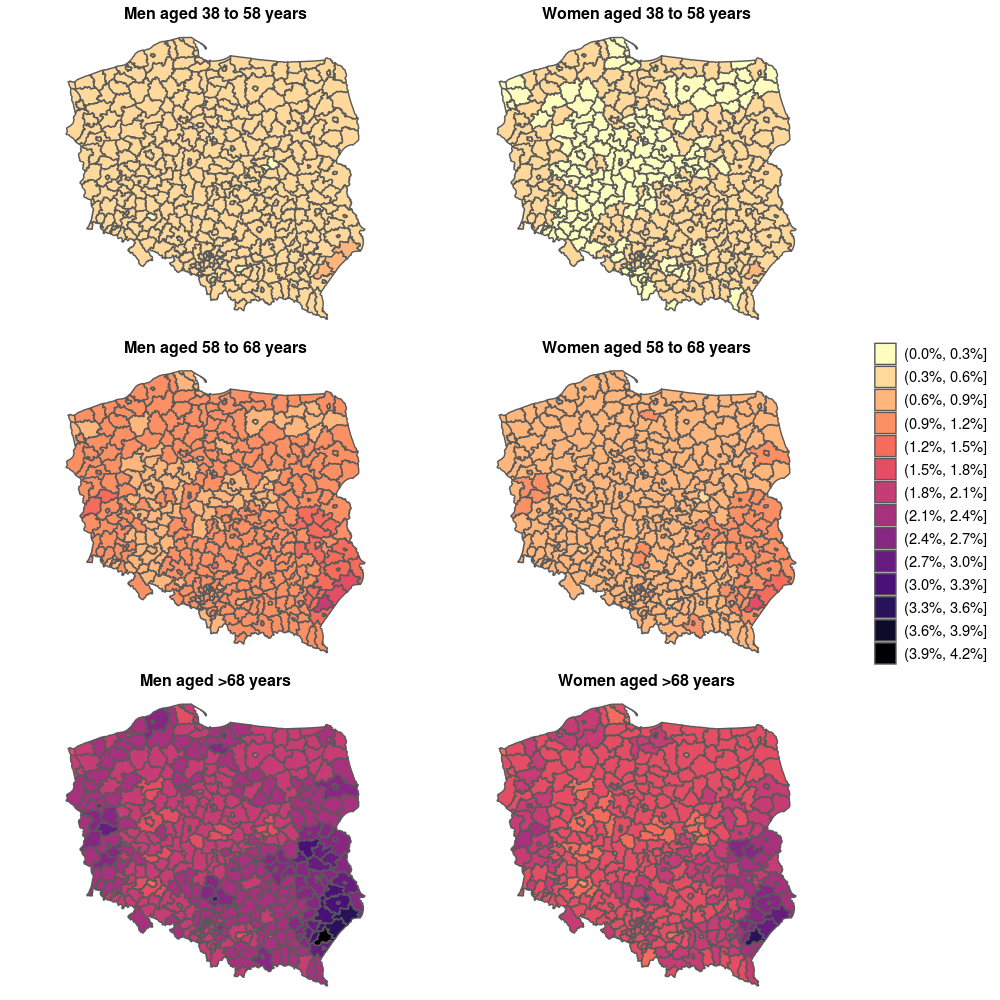}
\caption{Estimated probability of stroke by gender and age group based on the LOGIT SPATIAL model assuming that drugs for the cardiovascular system (ATC C) have been prescribed.}
\label{fig:results2}
\end{figure}

 The potential of the models analyzed is enormous because they allow us to study and visualise the outcomes of interest in relation to the  population subgroups   defined by the different values of the covariates.  This information is too long to be included in this article. By way of illustration we  present in Figure   \ref{fig:results} the posterior expectation  of the probability  of stroke, by gender and age group,  for people who did not take any medication, obtained from the spatial logistic regression model. It is clearly visible that the probability of stroke increases with age and in general women have lower probability than men. The largest difference between the estimated values is in the oldest age group. The spatial pattern is very relevant. In the southeast of Poland (Podkarpackie and Lubelskie Voivodeship) there is a visible spatial cluster with the highest risk of stroke. Among the ten powiats with the lowest estimated probabilities of stroke, nine of them are cities including Wroclaw, Cracow and the capital Warsaw. Similarly, and in accordance with the illustrative objective, Figure \ref{fig:results2} shows the posterior probabilities of stroke by
gender and age group  for people who takes drugs for the cardiovascular system (ATC C).
The overall pattern shows higher probabilities of stroke than in Figure
\ref{fig:results} due to the effect associated to these drugs (and the
underlying condition, i.e., cardiovascular diseases).

\section{Discussion}
\label{sec:discussion}

As previously stated in this paper, health care decisions often involve the collection
and analysis of datasets from different sources. Typical health data include
mortality and morbidity of certain diseases as well as other
information about risk factors, environmental exposure and others  \cite{health21st:2002}. In
addition, statistical  developments in recent years allow
researchers to handle, both methodologically and computationally, large datasets of individual data for population level
analyses that involve highly parameterized hierarchical models  \cite{bighealthdata:2014}.

Interesting analysis for health care decisions include the estimation of
prevalence, assessment of risk factors, estimation of spatial and temporal risk
variation, to mention a few. The assessment of risk factors is particularly
important because the identification (and prevention) of relevant risk factors
may help to reduce morbidity, which in turn may reduce mortality and public
health care expenditure.

Public health authorities can benefit from these population level analyses in
different ways. First of all, insight on a given condition can be gained by
conducting a population-wide analysis. Secondly, potential risk factors can be
assessed which can help to develop best health policies and practices. In the
study developed in this paper, a better understanding of the incidence of
stroke in Poland is gained as well as knowledge about potential risk factors,
with a particular interest on different conditions and associated prescription drugs. Given that health
care decisions by government agencies have an immediate and long-lasting effects
on the populations it is important that these decisions are data-driven.

In particular, this paper considers the analysis of population data about
stroke disease in Poland in a 2-year period. This is a
large dataset that comprises information about 500$\hspace*{0.04cm}$000 people on a number of
topics, including age, gender, other conditions and drugs prescribed, region and others. In
addition to the individual level data, information at the powiat level (such as
deprivation index and city/land county indicator) are available to complete the
analysis. Given the high burden of stroke, identifying risk factors which can
lead to a reduction in the prevalence of stroke will have a significant impact on the
overall quality of life of the population and the cost of public health care.

The available data can be approached in a number of different ways. First of
all, the probability of suffering from a stroke has been considered, for which
a logit analysis has been conducted. However, given that the time-to-stroke is
available, survival models can be used as well to tackle an
alternative inferential outcome. As individual and area level
data are available, multilevel models have been fitted. In addition to the
individual and area level covariates, mixed-effect models that include random
effects at the area level have been studied in two different ways: conditional independent and
spatially correlated random effects.

All these models have been estimated using a Bayesian framework, for which novel
computational methods have been used in order to fit the required models. In
particular, the integrated nested Laplace approximation \cite{Rueetal:2009}
has been used to obtain approximations of the posterior marginals of the parameters, random effects and
hyperparameters of the model. In addition, the implementation of INLA in the
R-INLA package is able to handle the hundreds of thousands of records in the dataset and
fit the models in a few minutes.

Relevant risk factors identified by the analysis include age, gender and certain conditions and associated drug prescriptions. In particular, women
showed a lower risk, which increased with age. Regarding the prescription drugs,
three different types of drugs (associated to relevant health risk factors of
stroke) were included in the models and they showed an increase in risk of
suffering from a stroke. However, our analysis is not able to disentangle whether this increased risk is due to the condition or the associated treatment. Furthermore, the estimates of both types of random
effects showed differences among powiats. Model selection using the DIC and
WAIC pointed to the model with fixed effects and spatially correlated random
effects as the best one among all the models proposed for both the logit and
survival families of models.

These models can be exploited for inference in a number of ways. The spatial
logit model can provide estimates of the probability of suffering a stroke for age, gender
and area. Similarly, survival models can provide estimates of time-to-stroke
for any individual or the median time-to-stroke according to age, gender and
area, and include the effect of prescription drugs in the estimates.

Other similar models can be used in the analysis of this dataset but the
proposed models provide additional opportunities for inference. As an example,
the output from the fitted models can be used for personalized medicine
provided that relevant individual-level information (e.g., genetic markers)
is available.

%%%%%%%%%%%%%%%%%%%%%%%%%%%%%%%%%%%%%%%%%%
\vspace{6pt}

%%%%%%%%%%%%%%%%%%%%%%%%%%%%%%%%%%%%%%%%%%
%% optional
%\supplementary{The following are available online at \linksupplementary{s1}, Figure S1: title, Table S1: title, Video S1: title.}


%%%%%%%%%%%%%%%%%%%%%%%%%%%%%%%%%%%%%%%%%%
\authorcontributions{Conceptualization, D.M., C.A. and V.G.; methodology, all authors; software, D.M. and V.G.; validation, all authors; formal analysis, all authors; investigation, all authors; resources, all authors; data curation, D.M.; writing--original draft preparation, all authors; writing--review and editing, all authors; visualization, D.M.; supervision, all authors; project administration, C.A., V.G. and P.P.; funding acquisition, C.A., V.G. and P.P. All authors have read and agreed to the published version of the manuscript.}
%'', please turn to the \href{http://img.mdpi.org/data/contributor-role-instruction.pdf}{CRediT taxonomy} for the term explanation. Authorship must be limited to those who have contributed substantially to the work reported.}

%%%%%%%%%%%%%%%%%%%%%%%%%%%%%%%%%%%%%%%%%%
\funding{This work was supported by the Project MECESBAYES
(SBPLY/17/180501/000491)
from the Consejer\'ia de Educaci\'on, Cultura y Deportes, Junta de Comunidades
de Castilla-La Mancha (Spain) and research grants
PID2019-106341GB-I00 and RTI2018-096072-B-I00
from Ministerio de Ciencia e Innovación (Spain).
D. M\l{}ynarczyk has been supported by a FPI research contract from
Ministerio de Ciencia e Innovación (Spain).}

%%%%%%%%%%%%%%%%%%%%%%%%%%%%%%%%%%%%%%%%%%
\acknowledgments{We would like to thank dr hab. Maciej Sm\k{e}tkowski for providing data on the deprivation index in Poland.}

%%%%%%%%%%%%%%%%%%%%%%%%%%%%%%%%%%%%%%%%%%
\conflictsofinterest{The authors declare no conflict of interest.}

%%%%%%%%%%%%%%%%%%%%%%%%%%%%%%%%%%%%%%%%%%
%% Only for journal Encyclopedia
%\entrylink{The Link to this entry published on the encyclopedia platform.}

%%%%%%%%%%%%%%%%%%%%%%%%%%%%%%%%%%%%%%%%%%
%% Optional
\abbreviations{The following abbreviations are used in this manuscript:\\

\noindent
\begin{tabular}{@{}ll}
AFT & Accelerated failure time\\
ATC & Anatomical Therapeutic Chemical \\
DIC & Deviance information criterion\\
ICAR & Intrinsic Conditional Auto-Regressive\\
INLA & Integrated nested Laplace approximation\\
GMRF & Gaussian Markov random field\\
MCMC & Markov chain Monte Carlo\\
PID & Powiat index of deprivation\\
WAIC & Watanabe-Akaike information criterion\\
\end{tabular}}

%%%%%%%%%%%%%%%%%%%%%%%%%%%%%%%%%%%%%%%%%%
%% Optional
\appendixtitles{no} % Leave argument "no" if all appendix headings stay EMPTY (then no dot is printed after "Appendix A"). If the appendix sections contain a heading then change the argument to "yes".
%\appendix
%\section{}
%\unskip
%\subsection{}
%The appendix is an optional section that can contain details and data supplemental to the main text. For example, explanations of experimental details that would disrupt the flow of the main text, but nonetheless remain crucial to understanding and reproducing the research shown; figures of replicates for experiments of which representative data is shown in the main text can be added here if brief, or as Supplementary data. Mathematical proofs of results not central to the paper can be added as an appendix.
%
%\section{}
%All appendix sections must be cited in the main text. In the appendixes, Figures, Tables, etc. should be labeled starting with `A', e.g., Figure A1, Figure A2, etc.
%

%%%%%%%%%%%%%%%%%%%%%%%%%%%%%%%%%%%%%%%%%%
\reftitle{References}


\externalbibliography{yes}
\bibliography{ms}

% The following MDPI journals use author-date citation: Arts, Econometrics, Economies, Genealogy, Humanities, IJFS, JRFM, Laws, Religions, Risks, Social Sciences. For those journals, please follow the formatting guidelines on http://www.mdpi.com/authors/references
% To cite two works by the same author: \citeauthor{ref-journal-1a} (\citeyear{ref-journal-1a}, \citeyear{ref-journal-1b}). This produces: Whittaker (1967, 1975)
% To cite two works by the same author with specific pages: \citeauthor{ref-journal-3a} (\citeyear{ref-journal-3a}, p. 328; \citeyear{ref-journal-3b}, p.475). This produces: Wong (1999, p. 328; 2000, p. 475)

%% for journal Sci
%\reviewreports{\\
%Reviewer 1 comments and authors’ response\\
%Reviewer 2 comments and authors’ response\\
%Reviewer 3 comments and authors’ response
%}

%%%%%%%%%%%%%%%%%%%%%%%%%%%%%%%%%%%%%%%%%%
\end{document}